# 45-year CPU evolution: one law and two equations

Daniel Etiemble
LRI-CNRS
University Paris Sud
Orsay, France
de@lri.fr

*Abstract—* **Moore's law and two equations allow to explain the main trends of CPU evolution since MOS technologies have been used to implement microprocessors.**

## I. INTRODUCTION

A new era started when MOS technologies were used to build microprocessors. After pMOS (Intel 4004 in 1971) and nMOS (Intel 8080 in 1974), CMOS became quickly the leading technology, used by Intel since 1985 with 80386 CPU.

MOS technologies obey an empirical law, stated in 1965 and known as Moore's law: the number of transistors integrated on a chip doubles every N months. Fig. 1 presents the evolution for DRAM memories, processors (MPU) and three types of read-only memories [1]. The growth rate decreases with years, from a doubling every 12 months to every 18 months then every 24 months. The semi-log scale clearly highlights the exponential growth of the number of transistors per chip. It turns out that this law and two basic equations allow to explain the main trends in the development of CPU and computer architectures since the mid's 70.

The CPU Performance Equation (1) is probably the most significant equation of the "quantitative approach" promoted in the different editions of Hennessy and Paterson's books [2].

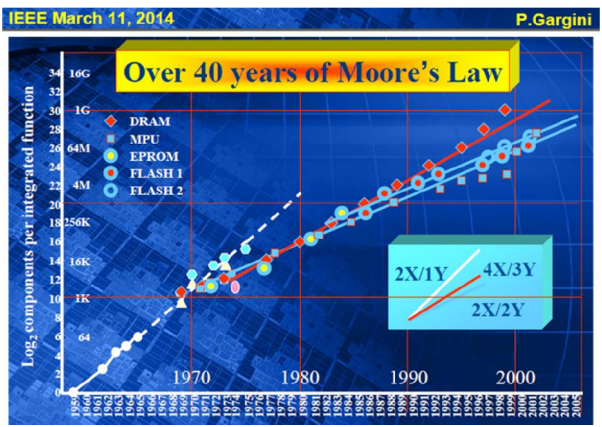

Fig. 1. Moore's Law until 2005

$$\text{CPU time} = IC \times CPI \times T_c = IC / (IPC \times F) \quad (1)$$

a) IC is the instruction count.
b) CPI is the clock cycles per instruction and IPC = 1/CPI is the Instruction count per clock cycle.
c) $T_c$ is the clock cycle time and $F=1/T_c$ is the clock frequency.

The Power dissipation of CMOS circuits is the second equation (2). CMOS power dissipation is decomposed into static and dynamic powers. For dynamic power, $V_{dd}$ is the power supply, F is the clock frequency, $\Sigma C_i$ is the sum of gate and interconnection capacitances and $\alpha$ is the average percentage of switching capacitances: $\alpha$ is the activity factor of the overall circuit

$$P_d = P_{d\,static} + \alpha \times \Sigma C_i \times V_{dd}^2 \times F \quad (2)$$

## II. CONSEQUENCES OF MOORE LAW

### A. Technological nodes

The increase of the number of transistors by chip results from the relatively regular launching of a new generation of CMOS technologies, called a technological node. The different successive nodes are shown in Fig. 2. The nodes were first defined according to the transistor channel length. The last nodes are more defined according to marketing criteria. However, every new node leads to reduced transistor sizes.

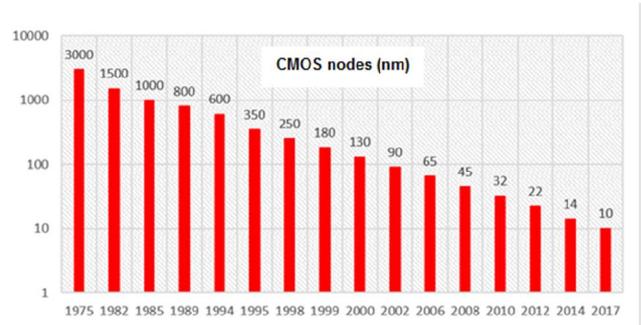

Fig. 2. CMOS technological nodes.

Moving from one technological node to the next one
a) Decreases the delays of logical gates and allows increased clock frequencies.
b) Increases the number of transistors per unit of area.

Discussing scaling rules is out of the scope of this paper. As a first approximation, the number of transistors per unit of area doubles and the gate delays are divided by 1.4 from one technological node to the next one, but with an increasing impact of the interconnection delays.

*B. Exponential growth and mismatch between exponentials*

Fig. 3 shows the increase in transistor count with the successive nodes. Figures are shown for three types of components: microprocessors, graphics processors (GPUs) and FPGAs. The numbers of transistors are those of the first components that use each node. The trend of the three curves is an exponential progression. Unfortunately, not all the characteristics of the DRAMs used for the main memories have the same growth rate, as shown in Fig. 4: if the surface of the memory cells depends directly on the dimensions of the nodes, the latency only decreased by 20% from 2000 to 2011.

All the exponential characteristics don't have the same growth rate, as shown in Fig. 5. The big issue is the mismatch between the CPU growth rate (clock frequency and performance) and the DRAM growth rates (bandwidth and latency) The huge difference between CPU and DRAM growth rates led to the increased complexity of microprocessor memory hierarchies. Different levels of caches are needed to balance the differences in the bandwidth and latency needs of the CPU with those of the DRAM main memory. Fig. 6 shows the evolution of cache hierarchies from Intel 80386 (1986) to IBM Power8 (2014):

  a) The number of cache levels increases: 1 then 2 then 3 then 4 cache levels. From the first to the last cache level, the cache capacity and the cache access time both increases.
  b) The larger cache is first in a different chip than the CPU chip and then integrated within the CPU chip. The single cache is external in the 80386. Then the L2 cache is external (Pentium). L3 for Power6 and L4 for the Power8 are located in a separated chip.

Disposing of more and more transistors on a chip allows to move from several chips to one chip for different features:
a) From external to internal caches.
b) From floating point coprocessors to integrated FP units.
c) From multiprocessors to multicores.
d) Etc.

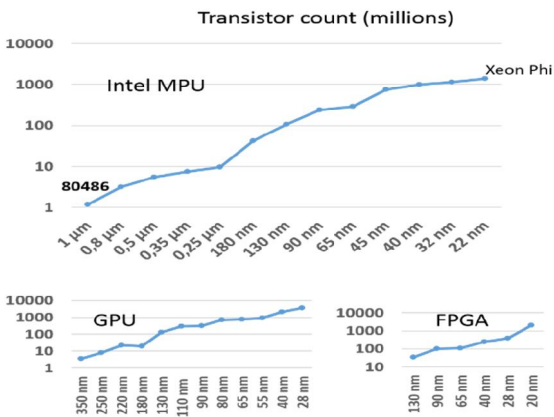

Fig. 3. Transistor count for the different nodes

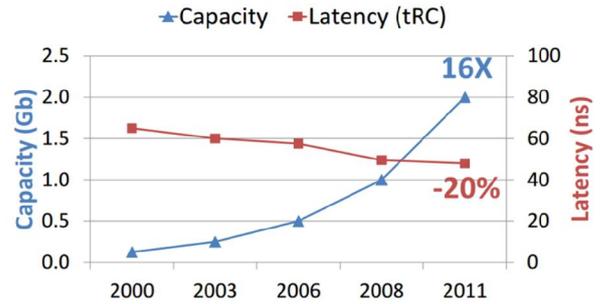

Fig. 4. Evolution of DRAM features

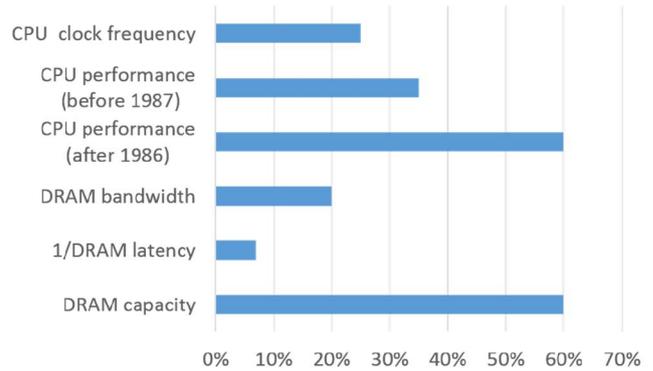

Fig. 5. Annual growth rate of different features

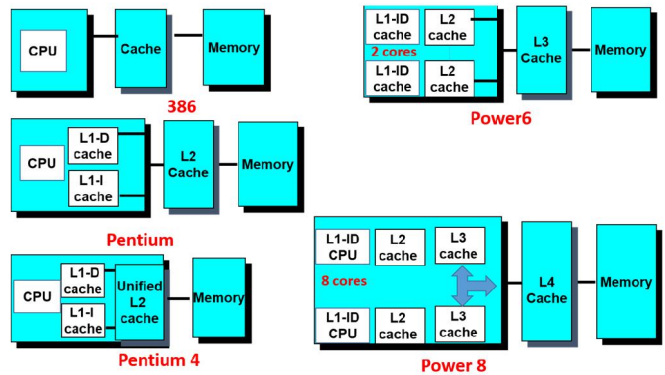

Fig. 6. Cache hierarchies from 1986 to 2014

The increased number of transistor also allows to add more and more features on a chip. On current SoC circuits can be found CPUs, DSP, GPU, Memory controllers, crypto components, specialized interfaces for the different standards of transmission, graphics, media, etc.

### III. CPU PERFORMANCE EQUATION

According to Equation (1), reducing the execution time of a program can be obtained by reducing either IC or CPI or Tc or both of them. Reducing CPI is increasing IPC and reducing Tc means increasing F. Since the first microprocessor, the relative impact of each term has evolved during the different periods and led to the major shift from mono-processors to multicore CPU with the "heat wall". While the different terms have evolved simultaneously, we present them separately for the clarity of the presentation.

### A. Increasing clock frequency

Increasing clock frequency is the simplest way to increase performance. The successive nodes of CMOS technologies lead to x1.4 decrease of the gate delays. It led to a 25% increase per year of clock frequencies from 740 kHz (Intel 4004) to 3 GHz (Intel Xeons with 45-nm nodes).

For a given node, performance is more than just Megahertz. By the end of 90's. A 200-MHz MIPS R10000 and a 400-MHz Alpha 21164 had approximately the same performance when running the same programs, while they differ by a factor up to 2 in clock speed. More, a 135-MHz Power2 had better FP performance than a 400-MHz 21164. It comes out that performance comes from CPI/F and not only F. Higher F can lead to higher CPI, and CPI could be quite different for Int and FP programs. Brainiac CPUs did more work during each clock cycle, while Clock Demon CPUs [3] used more pipeline stages that allow higher clock frequencies, each pipeline stage doing less work.

Without any change in micro-architectural features, moving from one node to the next one allows the frequency gain and the subsequent performance gain without any CPI change. Techniques to decrease CPI are discussed in the next section.

In 2017, with the only exception of the water cooled IBM z14 CPU (5.2 GHz), clock frequencies are limited in the 3-4 GHz range. However, 45-nm nodes (2008) provided 3-GHz clock frequencies. With 2017 10-nm technologies, it is quite obvious than tens of GHz would be available. The CPU frequency limitation is related to the "heat wall" [4] illustrated by Fig. 7. According to equation (2), the dynamic power dissipation is proportional to the clock frequency. The highest clock frequencies are limited just to keep the energy budget limits compatible with a safe operation of the components, and the typical IC packages and cooling techniques. This is the main reason for the shift towards multicore architectures, more cores replacing bigger cores, discussed in Section III-D, while intrinsic limitations of mono-processors is another reason.

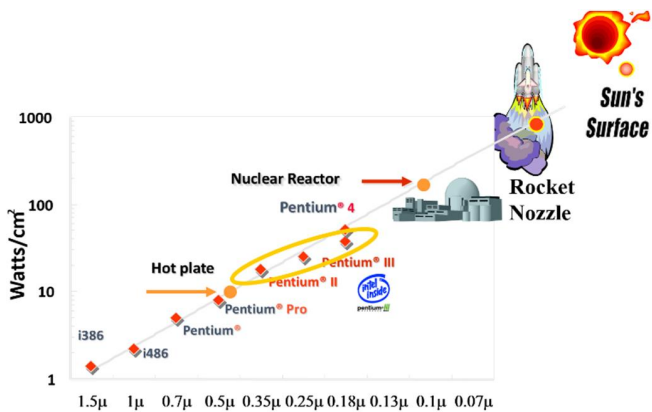

Fig. 7. Power density for successive nodes

### B. Reducing CPI and increasing IPC

CPI has two components:

a) The CPU CPI is the clock count per instruction assuming that there is no cycle during which the CPU is waiting for memory data. This component mainly relies on exploiting the instruction parallelism in sequential programs.
b) The Memory CPI is the clock count per instruction during which the CPU is waiting data from memory. This component mainly relies on the structure of the memory hierarchy.

*1) CPI and instruction parallelism*

Pipelining has been the first step to reduce the processor CPI. The most famous example is the MIPS R2000 5-stage pipeline for integer and memory instructions. It overlaps the execution of several successive instructions. The minimum value is CPI=1, which cannot be obtained due to data and control hazards. Superpipelining consists in pipelining some stages of the simple pipeline to increase the clock frequency up to the technology capability. However, more pipeline stages increase the impact of data hazards for memory loads and control hazards for taken branches. In both cases, integer multiplications and divisions and all FP operations complicate the pipeline operations and need software techniques such as loop unrolling to keep CPI close to the minimal value.

Superscalar execution consists in issuing several instructions per clock with different approaches:

a) In-order superscalar CPUs issue a group of instructions obeying defined rules per clock cycle. A first part fetches and decodes several instructions, and test resources and dependencies to form a group of instructions ready to be issued. From the group buffer, the instructions are launched in the different pipelines.
b) Out-of-order CPUs fetch instructions into an instruction buffer from which instructions are issued and executed according to the data flow. Instructions are retired in program order.
c) VLIW execution is another approach relying on the compiler. The CPU executes a pipeline of very long instruction words that consist of several normal instructions. The compiler uses specific techniques such as loop pipelining to generate the VLIW instructions.

Out-of-order superscalar CPUs are the most efficient mono-processors, but using instruction parallelism has quickly been faced to the law of "diminishing return", as shown in Fig. 8. Intel (and AMD) superscalar CPUs decompose IA-32 and Intel64 CISC instructions into μops, which are simpler instructions similar to the RISC instructions of RISC ISAs. For Intel CPUs, from 1995 to 2013, the hardware resources (physical registers, ROB, reservation stations) have grown significantly, while the maximal μop count per cycle has only increased from 3 to 4. The additional resources are used to enlarge the number of μops considered for simultaneous launching without changing the maximum number of μops per clock. Obviously, the difficulties to extract instruction level parallelism from sequential programs also contribute to the IPC limits of the mono-processors.

*2) CPI and memory wall*

To reduce the memory CPI component, the cache hierarchy has evolved towards a larger number of levels, as it has been shown in Fig. 6. Similarly, the DRAM structures and interfaces have evolved to avoid a "memory wall". However, even with a reduced memory CPI component, pipeline stalls due to memory waits still exist when executing a single program.

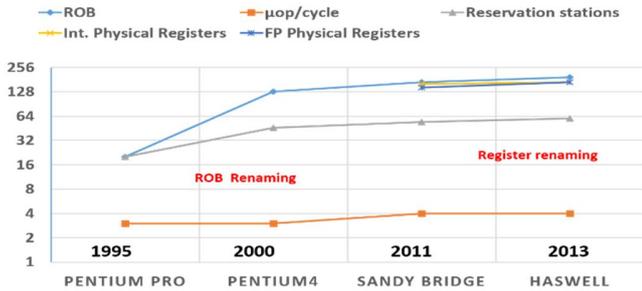

Fig. 8. IPC limits in Intel out-of-order CPUs

Memory CPI can be further reduced at the CPU level. Equation 1 can be used to determine the execution time of multithreaded execution of several programs on a CPU. There is no change in clock frequency and the instruction count is now the sum of the instruction count of the different programs. However, the overall CPI for these programs can be reduced with a hardware support for multithreading. The physical CPU has different architectural states (PC, architectural and state registers), several functional units and a cache hierarchy. There are several logical processors that share the functional units and the memory hierarchy. There is no reduction of the execution time of each individual program, but increase in the number of instructions from different programs executed per time unit.

Two types of multithreaded approaches have been used more and more since the early 2000s.

a) Fine grain multithreading: the CPU switches from one thread to another one in one clock cycle, either when there are pipeline hazards (multi-cycle operations or cache misses) or according to an algorithm for a fair distribution of resources to the different threads. Sun Niagara architecture [5] or the Oracle Sparc servers are typical examples of this approach. Overall CPI is reduced by suppressing most hazards in the different pipelines.

b) Simultaneous multithreading is used with out-of-order superscalar CPUs. Instructions from different threads are issued at each clock and executed. Intel hyperthreading with 2 threads [6] and IBM Power CPUs with 2 to 8 threads are typical of this approach. Again, overall CPI is reduced by suppressing hazards that would occur in individual execution of each thread.

C. *Decreasing IC*

For mono-processors, until mid90', the instruction count IC was determined by the execution of the code generated by the compiler. The "fight" between RISC and CISC instruction sets, mainly x86, focused on the IC x CPI product. For RISC ISA, IC was higher, but CPI was lower. With the dynamic translation of x86 instructions into RISC-like instructions called µops used from the Pentium Pro (1995), there are no significant differences between the numbers of executed µops or instructions (IC) for the different instruction sets.

Until mid90', the only way to significantly reduce IC was to use a parallel architecture, for which the instruction count is allocate to the different processors using either data parallelism or control (thread) parallelism. The two main types of parallel architectures are shown in Fig. 9. Left part shows the shared memory architectures (multiprocessors) for which Pthread or OpenMP are the most commonly used programming models. The right part of the figure shows the distributed memory architectures (multi-computers) for which message passing is used to communicate between CPUs. MPI is the commonly used programming interface for this type of architecture. Clusters can combine the two approaches. Except for simple cases, the dispatch of instructions to the different cores is not trivial. As soon as the architecture includes clusters, communication times must be considered. However, the IC of the most loaded processor is reduced compared to the overall IC.

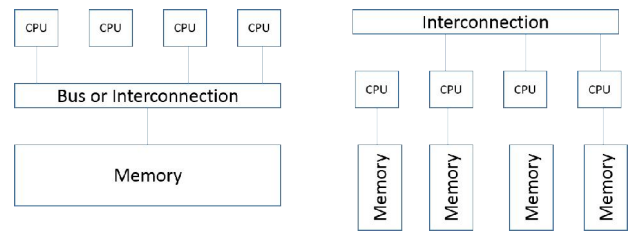

Fig. 9. Multiprocessors (left) and multi-computers (right)

Due to the exponential progresses close to 60% per year of mono-processors, until the "heat wall", using parallel architectures was limited to multiprocessors for servers and more complex parallel architectures for supercomputers. For "main stream" computing, it was the "free lunch", the golden age of scaling of mono-processors. For most applications, the best and cheapest way to increase performance was to wait for the next generation of mono-processors, which explains why parallel architectures were limited to an important, but limited niche until 2000'.

More, the integration capabilities of mid90' allowed to introduce data parallelism in mono-processors to reduce IC. This data parallelism has been used in two types of processors, as shown in Fig. 10.

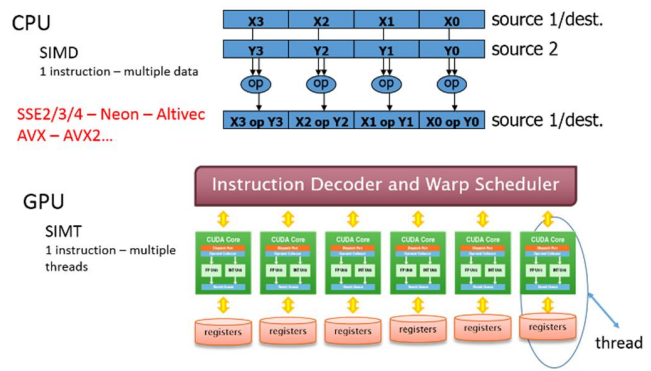

Fig. 10. SIMD and SIMT data parallelism

In general purpose CPUs, SIMD instructions execute the same operation on several data (higher part of Fig. 10) reducing the number of arithmetical and logical operations by a factor equal to the number of data per instruction. These SIMD instructions are available in nearly every ISA, from MMX with 64-bit registers to SSE (128-bit), AVX and AVX2 (256-bit), AVX512 (512-bit) for Intel64 instruction set. There exists for ARM (Neon variants), IBM Power, etc. SIMD instructions support 8-16-32 integer types according to SIMD register size and simple and double precision floating point numbers.

The graphics processors (GPU) launched by the end of 90' use the SIMT model (Single Instruction Multiple Threads) where one SIMT instruction execute the same instruction in several threads (lower part of Fig. 10). However, GPU programming model is different and uses programming language such as CUDA (NVidia) and OpenCL.

### D. The shift towards parallelism

For mono-processors, increased clock frequencies could no longer be used to increase performance ("heat wall") and instruction parallelism (ILP) in sequential programs has reached a plateau. The only possibility was to shift to multicore processors, which are new versions of multiprocessors for which all cores are integrated in the same chip. Fig. 11 compares the multicore processors with multithreaded processors and multiprocessors.

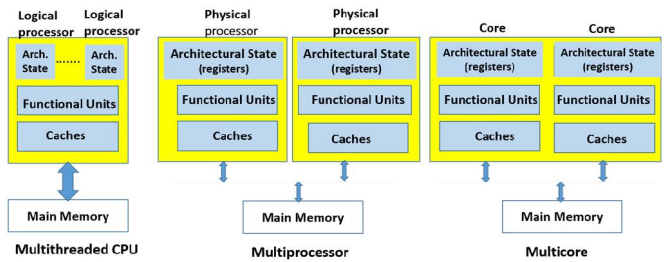

Fig. 11. Multithreaded, multiprocessor and multicore CPUs

#### 1) Energy consumption

Fig. 12 [7] compares the performance-power for mono-processors and multicores. Left part of the figure compares performance of a mono-processor at frequency $F_{max}$, with performance of the same processor at frequency 1.2 $F_{max}$ and performance of a dual-core at frequency 0.8 $F_{max}$. With a higher clock frequency, the mono-processor has a minimal increase on performance, but a significant increase of power consumption. On the other side, the dual core with a slightly decreased clock frequency has a small increase in power consumption, but a significant performance increase. The right part of the figure compares a small core, a large core (4x larger than the small one) and a multi-core with four small cores. As a first approximation, power consumption is proportional to the used area, while performance is proportional to $\sqrt{area}$. The 4-core has 2x the performance of the 1-core with the same power consumption. The 4-core and the 1-core have the same performance/power ratio.

#### 2) Intrinsic limits of mono-processors

The limited ILP in sequential programs first led to the introduction of multithreaded mono-processors, moving from sequential to parallel programming. However, the threads share the same set of execution units that limit performance. Moving to independent processors (core), possibly multithreaded, was the next step.

As the cores of multicores are implemented by superscalar (multithreaded or not) mono-processors, reducing IC by increasing the number of cores is the main technique for getting increased performance.

Shifting from monothreaded mono-processors to multi-cores also means shifting from sequential to parallel programming. You are now faced with Amdahl's law… and all the issues of parallel programming. This was outlined by D. Paterson in a famous paper [8]: "The Trouble with Multicore: Chipmakers are busy designing microprocessors that most programmers can't handle"

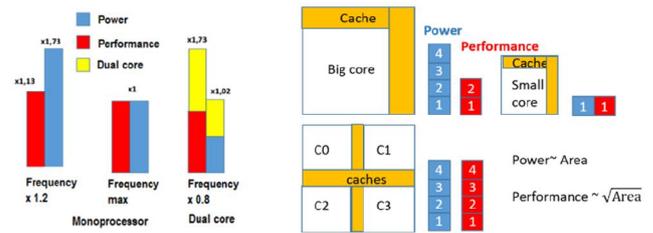

Fig. 12. Comparing mono-processors and multicores

### IV. CMOS POWER DISSIPATION

Old CMOS circuitry had no static power consumption, but situation has changed starting with 90-nm nodes due to leakage currents. $V_{dd}$ power supply has been reduced for a long time to the minimum value that allows the transistors to operate with correct "on" and "off" states. This minimum value is in the range of 0.8 to 1V. Until hitting the "heat wall", CMOS power consumption was mainly due to the clock frequency on one hand, and to $\alpha \sum C_i$ that increases with the number of transistors and interconnections that the successive technology nodes allow. Since the "heat wall", power consumption has become the major issue for which all components of Equation 2 must be considered.

### A. Reducing static power dissipation.

Technological improvements allow to reduce static power. For instance, the "off" current of the Intel tri-gate transistors is $10^{-5}$ the "on" current instead of $10^{-4}$ for a planar transistor.

Circuitry techniques can also be used to reduce static power. For instance, the L3 cache of the Xeon CPU [9] is decomposed into different sub-blocks that have different operation modes by using virtual ground levels. Compared to the active mode, there was x2 and x4 reduction of leakage currents with sleep and switch-off modes.

### B. Reducing the circuit activity

Reducing the dynamic power means reducing the circuit activity via $\alpha$, $V_{dd}$ and F parameters.

#### 1) Vdd and clock controls

In operating mode, $V_{dd}$ and clock values are bounded. But a clock can be stopped and power supplies can be disconnected by decomposing the chip floor plans into domains in which power supplies and clocks are controlled. With such domains, several power supply values and several clock frequencies can be defined and used, according to the circuit needs.

*2) Several operation modes*

Defining several operation modes to only activate the used parts of a circuit is largely used to reduce the power consumption. For instance, the 5th generation Intel Core M CPUs have different modes and sub-modes, from the "working" mode to the "mechanical off" mode. In complex microcontrollers, different blocks implementing different features have different operation modes, from the active mode down to other modes such as sleep, deep sleep, stop and shut-off.

## V. WHAT ABOUT FUTURE?

### A. Moore's law

The described evolution is based on Moore's law, i.e. an exponential function. As quoted by G. Moore in 2003, *"No Exponential is Forever: But "Forever" Can Be Delayed"*. This quote raises two questions: the end of the exponential and delaying the end. P. Gardini, Intel Fellow and chairman of ITRS [1] gives some insights.

a) First lesson: "*Predictors of Engineering Limits have Always been Proven Wrong by The Right Improvements*". Intel 3D tri-gate transistors or FinFET technologies are good examples of technical improvements that delay the limits.
b) Second lesson: "*It Would be Wrong to Believe that the Right Fundamental Limits Don't Exist*". Moore's law is now "slowing", as the fundamental limits are closer than ever.

### B. The execution time equation

Significant changes are doubtful for clock frequencies. CPI could be further decreased if new PIM architectures with DRAM-implemented CPUs become successful. Reducing the number N of executed instructions per core is still going on and will continue.

a) SIMD vector length has regularly increased and AVX-1024 is forecasted. On the other side, 16-bit FP SIMD computation, suggested in 2005 [10], is now used in recent NVidia GPUs and some ARM architectures.
b) While SIMD and SIMT introduced "vector" instruction types, recent processors or coprocessors focusing on AI and deep neural networks introduces 2D or "matrix" instructions and computations. Tensor cores in NVidia Volta [11], Matrix-Multiplication Unit in Google TPU [12] and Intelligent Processing Unit [13] (Graphcore) are good examples of the evolution towards pipelined CISC instructions, with increased CPI, while reducing IC.
c) Increasing the number of cores in multi and many-cores architectures reduces the IC of each core. Again, the number of cores regularly increases, but one can hardly predicts an exponential growth and unsolved software issues can prevent to fully exploit these many-cores architectures or restrict them to special applications.

### C. The power dissipation equation

It will be there as long as CMOS will be there.

## VI. CONCLUDING REMARKS

Since using CMOS technology, Moore's Law has been continuously valid, even though the period of doubling the number of transistors has grown. The exponential growth is there, even if DRAM features lead to more and more complex memory hierarchies.

During these years, the equation of a program execution time is an insight to understand the factors that have been used to increase performance: first, increasing clock frequencies until hitting the "heat wall", with the additional use of pipelining and instruction parallelism, SIMD and SIMT data parallelism in mono-processors. The exponential increase of performance in main stream computing has limited parallel architectures to servers and supercomputers. Then came the period of parallel architectures as multicores and manycores are the main technique to continue performance increase. As "heat wall" has become the big issue, the equation for CMOS power dissipation is important to explain the different techniques that have been used to limit power dissipation.

Moore's law and two equations have been able to explain the evolution of Von Neumann architectures for more than 40 years. When the end of the predicted end of the exponential evolution will be real or when non-Von Neumann architectures will prove to be more efficient for programmable applications, the situation will be totally different. Until that point, the two equations that have been discussed in this paper will be there to explain the evolution.